\documentclass[aps,prx,superscriptaddress,twocolumn,nofootinbib,nobibnotes,floatfix,showpacs,reprint,longbibliography]{revtex4-2}
\usepackage{siunitx}
\usepackage[utf8]{inputenc}
\usepackage[T1]{fontenc}
\usepackage{lmodern}
\usepackage{orcidlink}
\usepackage{amsmath,amsfonts,amssymb,amsthm}
\usepackage{bm} % Bold math
\usepackage{xcolor} % For text color
\usepackage{xfrac} % For slant fractions \sfrac{}{}
\usepackage{enumitem} % Nested enumeration
\usepackage[normalem]{ulem} % For underlining and emphasis, may cause problems with citation line-breaks
\usepackage{soul} % For highlighting and striking out text
\usepackage{subfigure}
\usepackage{multirow}
\usepackage{dcolumn} % Align table columns on decimal point
\usepackage{graphicx} % Required for inserting images
\usepackage{hyperref} % Add hypertext capabilities
\hypersetup{
    unicode={true},
    colorlinks={true},
    linkcolor={blue},
    citecolor={blue},
    urlcolor={blue}
}

\begin{document}
\title{Lead-free antiperovskite derivatives Ba$_{3}$MA$_{3}$ (M = P, As,
Sb, Bi; A = Cl, Br, I): Next-gen materials for optoelectronics}
\author{Surajit Adhikari}
\email{sa731@snu.edu.in}

\affiliation{Department of Physics, Indian Institute of Technology Bombay, Powai,
Mumbai 400076, India}

\affiliation{Department of Physics, School of Natural Sciences, Shiv Nadar Institution
of Eminence, Greater Noida, Gautam Buddha Nagar, Uttar Pradesh 201314,
India}

\author{Aftab Alam}
\email{aftab@iitb.ac.in}

\affiliation{Department of Physics, Indian Institute of Technology Bombay, Powai, Mumbai 400076, India}
\affiliation{Centre for Machine Intelligence and Data Science (CMInDS), Indian Institute of Technology Bombay, Powai, Mumbai 400076, India}

\author{Priya Johari}
\email{priya.johari@snu.edu.in}

\affiliation{Department of Physics, School of Natural Sciences, Shiv Nadar Institution
of Eminence, Greater Noida, Gautam Buddha Nagar, Uttar Pradesh 201314,
India}
\begin{abstract}
Antiperovskite derivatives have recently emerged as promising lead-free alternatives to halide perovskites for optoelectronic applications. Here, using a comprehensive first-principles calculations including density functional perturbation theory and many-body perturbation theory (involving GW and Bethe-Salpeter equation (BSE)),  we investigate the stability, excitonic, polaronic, and optoelectronic properties of cubic Ba$_3$MA$_3$ (M = P, As, Sb, Bi; A = Cl, Br, I). These compounds are found to be dynamically and thermodynamically stable direct-gap semiconductors with G$_0$W$_0$@PBE+SOC band gaps spanning 1.23--2.17 eV. BSE calculations reveal moderate exciton binding energies (0.254--0.352 eV) and intermediate-radius excitons, while Fr\"ohlich polaron analysis indicates intermediate carrier--phonon coupling and mobilities up to $\sim$ 75 cm$^{2}$V$^{-1}$s$^{-1}$. The resulting spectroscopic limited maximum efficiencies reach $\sim$ 19--32\%, surpassing several lead-based perovskites. Our results establish Ba-based antiperovskite derivatives as a robust, eco-friendly platform for next-generation optoelectronic devices.
\end{abstract}

\maketitle

{\it Introduction:}
Lead halide perovskites have attracted intense research interest owing to their exceptional optoelectronic properties, with power conversion efficiencies (PCEs) increasing from 3.8\% to over 27\% within a decade \citep{chapter2-12,chapter3-8,chapter3-10,chapter2-53}. However, the toxicity due to lead and limited long-term stability impede their commercial deployment \citep{chapter3-12,chapter1-16}, motivating the search for safer and more stable alternatives. In this context, antiperovskite semiconductors (X$_{3}$BA), which are structurally analogous to conventional perovskites (ABX$_{3}$) but feature reversed cation–anion roles, have emerged as promising candidates for optoelectronic applications \citep{AP-1,AP-2}. Ion inversion in antiperovskites produces distinct band-edge characteristics compared to halide perovskites, enabling a broader compositional design space.

Further development has led to antiperovskite derivatives (X$_{3}$BA$_{3}$), obtained through structural modifications such as anion ordering and splitting. These retain the three-dimensional octahedral framework while offering enhanced thermodynamic stability and tunable bandgaps \citep{AP-4,AP-7,AP-8}. Notably, several members of this family exhibit optoelectronic properties and theoretical solar conversion efficiencies comparable to, or exceeding, those of lead-based perovskites, highlighting their potential as lead-free, high-performance semiconductors \citep{AP-4,AP-5,AP-6,AP-7,AP-8}. In X$_{3}$BA$_{3}$ compounds, alkaline-earth metals (Mg, Ca, Sr, Ba) occupy the X-site, pnictogens (N, P, As, Sb, Bi) the B-site, and halides (F, Cl, Br, I) the A-site. The pnictogen–halide combination is particularly favorable for achieving direct bandgaps, strong optical absorption, and structural stability.

Experimentally, Mg$_{3}$NF$_{3}$ has been synthesized at high temperatures (1050~$^\circ$C) and adopts a cubic Pm$\bar{3}$m structure \citep{AP-10}. While numerous theoretical studies have investigated inorganic antiperovskite derivatives \citep{AP-3,AP-4,AP-5,AP-7,AP-8,AP-11}, Ba-based systems have received special attention due to their larger ionic radii, lattice flexibility, and direct bandgaps suitable for photovoltaic applications \citep{AP-4,AP-8,AP-12}. Despite their promising photovoltaic performance, the excitonic and polaronic properties of these materials remain largely unexplored, even though they critically influence charge separation, carrier mobility, recombination, and realistic device efficiency.

In this work, we present the first comprehensive study of excitonic and polaronic physics in antiperovskite derivatives Ba$_{3}$MA$_{3}$ (M = P, As, Sb, Bi; A = Cl, Br, I). Many-body effects and carrier--phonon interactions are explicitly examined together with electronic and optical properties and theoretical efficiency. Our computational approach combines density functional theory (DFT) \citep{chapter2-36,chapter2-37}, density functional perturbation theory (DFPT) \citep{chapter1-60}, and many-body perturbation theory (MBPT) \citep{chapter3-1,chapter3-2} to evaluate electronic structure, dielectric screening, exciton binding energies, carrier--phonon coupling, polaron mobilities, and spectroscopic limited maximum efficiency (SLME). All calculations are performed using the Vienna \textit{Ab initio} Simulation Package (VASP) \citep{chapter1-31,chapter1-32}. Additional computational details and supporting results are provided in Sec. I of the Supplemental Material (SM) \citep{supp}; see also Refs.~\citep{chapter1-33,chapter2-3,chapter3-6,chapter2-10,chapter1-47,chapter1-48,chapter1-49,chapter1-50,chapter1-73,chapter1-74,chapter1-75}.

\begin{figure}[t]
\begin{centering}
\includegraphics[width=0.5\textwidth,height=0.75\textheight,keepaspectratio]{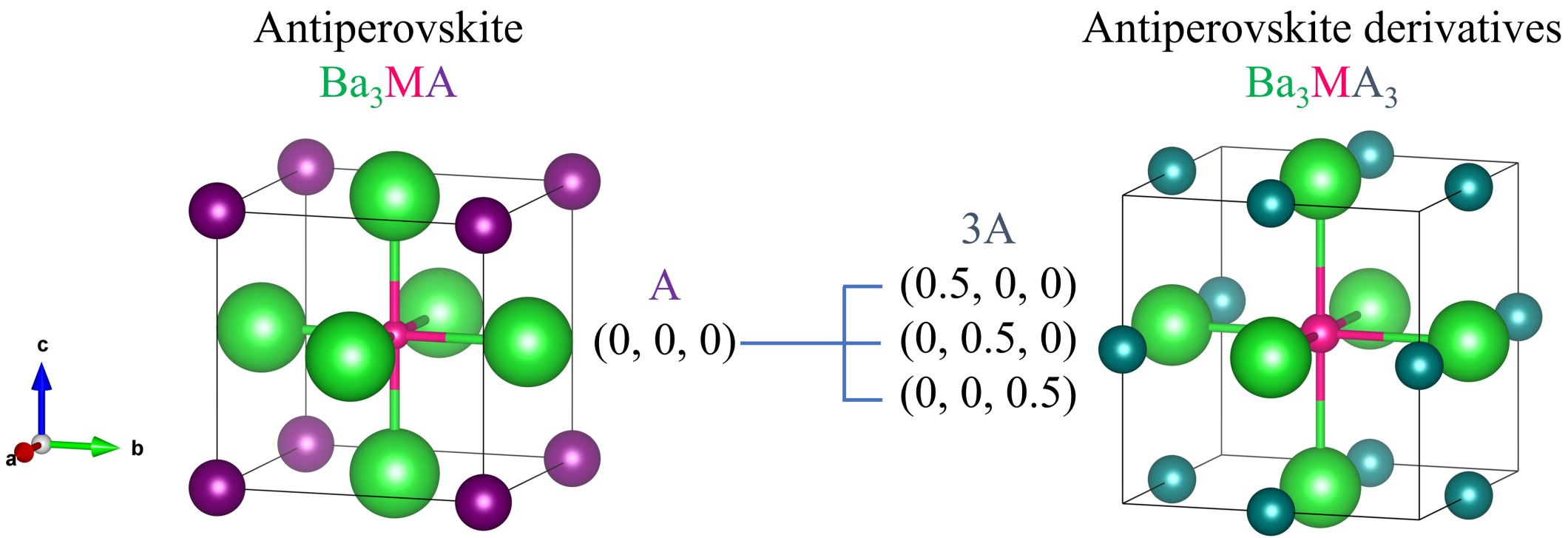}
\par\end{centering}
\caption{\label{fig:1}Design strategy for antiperovskite derivatives obtained by splitting the A-site position at (0,0,0) into three distinct positions (0.5,0,0), (0,0.5,0), and (0,0,0.5).}
\end{figure}

{\it Results and Discussions:}
Structurally, antiperovskites (X$_{3}$MA$_{3}$) are derived from conventional perovskites (AMX$_{3}$) via cation--anion inversion while preserving symmetry. Their derivatives further modify this framework by replacing one trivalent anion (A$^{3-}$) with three monovalent anions (3$\times$A$^{-}$) to maintain charge neutrality (Fig.~\ref{fig:1}), which can be viewed as successive cation--anion inversion and A-site splitting \citep{AP-9}. Figure~\ref{fig:1}(b) shows the resulting Ba$_{3}$MA$_{3}$ antiperovskite derivatives (M$^{3-}$ = P, As, Sb, Bi; A$^{-}$ = Cl, Br, I). We focus on the cubic Pm$\bar{3}$m (No.~221) phase, which combines features of halide perovskites and antiperovskites and is suitable for optoelectronic applications. In this structure, each Ba$^{2+}$ is octahedrally coordinated by two M$^{3-}$ and four A$^{-}$ anions, yielding equivalent Ba--M and Ba--A bond lengths (Table~S1). The PBE-calculated lattice parameters (Table~\ref{tab:1}) agree with previous studies \citep{AP-3,AP-5,AP-9}. Lattice constants increase from Cl $\rightarrow$ Br $\rightarrow$ I for fixed M and from P $\rightarrow$ As $\rightarrow$ Sb $\rightarrow$ Bi for fixed A, consistent with bond-length trends. Structural stability is confirmed by phonon dispersion \citep{chapter1-60} (Fig.~S1), as well as decomposition enthalpies and elastic constants, indicating thermodynamic and mechanical stability (see Secs. IV and V of the SM \cite{supp}).

\begin{table*}[t]
\caption{\label{tab:1}Calculated lattice parameters (in $\textrm{\AA}$), electronic band gaps (in eV) from different methods, effective electron ($m_{e}^{*}$) and hole ($m_{h}^{*}$) masses, and reduced masses ($\mu^{*}$) (in units of $m_{0}$) for Ba$_{3}$MA$_{3}$ (M = P, As, Sb, Bi; A = Cl, Br, I).}

\begin{centering}
\begin{tabular}{ccccccccccccc}
\hline 
\multirow{2}{*}{Systems} & \multirow{2}{*}{} & \multirow{2}{*}{A} & \multirow{2}{*}{} & \multicolumn{2}{c}{Lattice parameters (a = b = c)} & \multirow{2}{*}{} & \multicolumn{2}{c}{Bandgap (eV)} & \multirow{2}{*}{} & \multirow{2}{*}{$m_{e}^{*}$ ($m_{0}$)} & \multirow{2}{*}{$m_{h}^{*}$ ($m_{0}$)} & \multirow{2}{*}{$\mu^{*}$ ($m_{0}$)}\tabularnewline
\cline{5-6} \cline{6-6} \cline{8-9} \cline{9-9} 
 &  &  &  & This work & Previous reports &  & HSE06+SOC & G$_{0}$W$_{0}$@PBE+SOC &  &  &  & \tabularnewline
\hline 
\multirow{3}{*}{Ba$_{3}$PA$_{3}$} &  & Cl &  & 6.44 & 6.42 \citep{AP-3} &  & 1.64 & 2.17 &  & 0.705 & 0.380 & 0.247\tabularnewline
 &  & Br &  & 6.62 & 6.61 \citep{AP-5} &  & 1.60 & 2.05 &  & 0.594 & 0.417 & 0.245\tabularnewline
 &  & I &  & 6.87 &  &  & 1.38 & 1.72 &  & 0.453 & 0.496 & 0.237\tabularnewline
 &  &  &  &  &  &  &  &  &  &  &  & \tabularnewline
\multirow{3}{*}{Ba$_{3}$AsA$_{3}$} &  & Cl &  & 6.51 &  &  & 1.60 & 2.04 &  & 0.755 & 0.432 & 0.275\tabularnewline
 &  & Br &  & 6.66 & 6.67 \citep{AP-5} &  & 1.52 & 1.91 &  & 0.645 & 0.436 & 0.260\tabularnewline
 &  & I &  & 6.91 &  &  & 1.29 & 1.57 &  & 0.470 & 0.423 & 0.223\tabularnewline
 &  &  &  &  &  &  &  &  &  &  &  & \tabularnewline
\multirow{3}{*}{Ba$_{3}$SbA$_{3}$} &  & Cl &  & 6.70 & 6.68 \citep{AP-3} &  & 1.55 & 1.98 &  & 0.691 & 0.360 & 0.237\tabularnewline
 &  & Br &  & 6.85 & 6.83 \citep{AP-5} &  & 1.48 & 1.86 &  & 0.711 & 0.363 & 0.240\tabularnewline
 &  & I &  & 7.11 &  &  & 1.30 & 1.60 &  & 0.582 & 0.362 & 0.223\tabularnewline
 &  &  &  &  &  &  &  &  &  &  &  & \tabularnewline
\multirow{3}{*}{Ba$_{3}$BiA$_{3}$} &  & Cl &  & 6.74 &  &  & 1.13 & 1.64 &  & 0.520 & 0.255 & 0.171\tabularnewline
 &  & Br &  & 6.87 & 6.88 \citep{AP-9} &  & 1.05 & 1.49 &  & 0.526 & 0.249 & 0.169\tabularnewline
 &  & I &  & 7.11 & 7.11 \citep{AP-9} &  & 0.87 & 1.23 &  & 0.597 & 0.220 & 0.161\tabularnewline
\hline 
\end{tabular}
\par\end{centering}
\end{table*}

\begin{figure*}[t]
\includegraphics[width=0.8\textwidth,height=1\textheight,keepaspectratio]{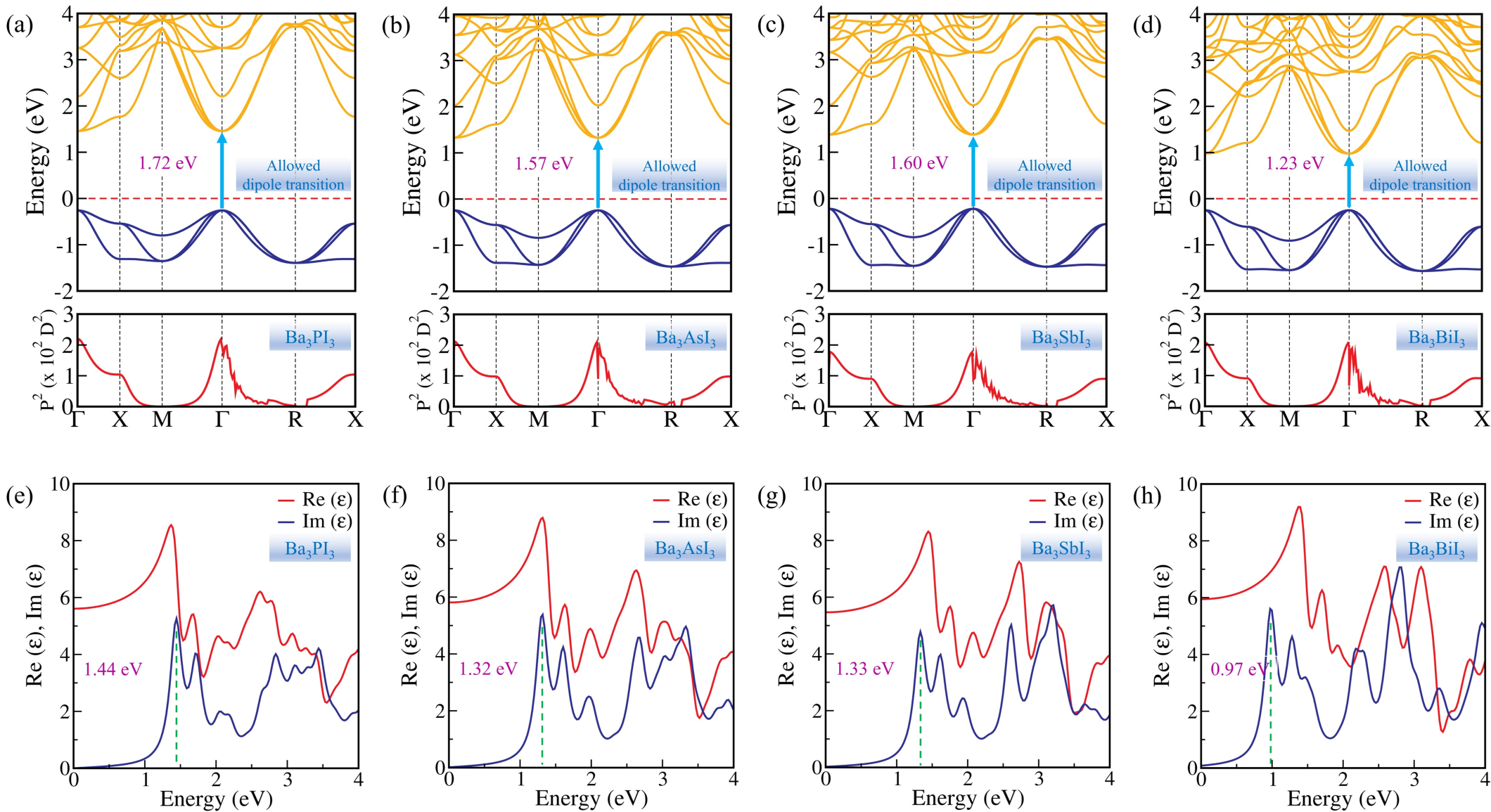}
\caption{(a-d) Electronic band structures and transition probabilities calculated using G$_{0}$W$_{0}$@PBE+SOC and (e-h) real and imaginary parts of the dielectric function calculated using BSE@G$_{0}$W$_{0}$@PBE+SOC for Ba$_{3}$MI$_{3}$ (M = P, As, Sb, Bi).}
\label{fig:2}
\end{figure*}

After establishing structural stability, we examine the electronic properties of Ba$_{3}$MA$_{3}$ using density of states and band-structure calculations. Atom- and symmetry-projected DOS near the band edges are shown in Fig.~S2 of the SM \cite{supp}. Band gaps obtained with the PBE functional \citep{chapter1-34} are underestimated due to self-interaction errors (Table~S4). More accurate results are achieved using the HSE06 hybrid functional \citep{chapter1-35} and the G$_{0}$W$_{0}$@PBE approach \citep{chapter1-69,chapter1-70}. Spin--orbit coupling (SOC) is included in PBE and HSE06 calculations and is particularly important for Sb- and Bi-based compounds. \textcolor{black}{Fully self-consistent G$_{0}$W$_{0}$@PBE+SOC calculations are computationally very demanding for the considered systems (relatively large unit cell size) and convergence settings. Therefore, SOC effects are incorporated perturbatively by adding the SOC-corrected band-gaps obtained from HSE06 calculations to the scalar-relativistic G$_{0}$W$_{0}$@PBE band gaps when plotting the band structures.} This approach is justified since the band-gap difference ($\Delta E_g$) with and without SOC remains nearly identical across functionals for all compounds (Table~S4, SM \cite{supp}). Figures~S5 and Fig.~\ref{fig:2}(a--d) show that all systems are direct-bandgap semiconductors with both VBM and CBM at the $\Gamma$ point. The HSE06 band gaps of Ba$_{3}$MBr$_{3}$ (M = P, As, Sb) agree with earlier reports \citep{AP-5}. The resulting G$_{0}$W$_{0}$@PBE+SOC band gaps (E$_g$) range from 1.23 to 2.17~eV, comparable to lead halide perovskites (1.50--1.85~eV) \citep{chapter5-11,chapter5-13}, highlighting their potential for photovoltaic and optoelectronic applications. The calculated electron and hole effective masses ($m_e^{*}$, $m_h^{*}$; Table~\ref{tab:1}) further indicate predominantly p-type transport behavior (see Sec. IX of the SM \cite{supp}).

Although these compounds exhibit favorable band gaps and low carrier effective masses, these features alone do not establish their suitability as photovoltaic absorbers. A detailed analysis of their optical properties, particularly the frequency-dependent dielectric function $\varepsilon(\omega)$, is therefore required. To improve accuracy, we employ MBPT-based BSE@G$_{0}$W$_{0}$@PBE calculations that explicitly include electron--hole interactions \citep{chapter1-67,chapter1-68}. As in the band-structure analysis, SOC-corrected $E_g$ values obtained at the HSE06 level are incorporated into the BSE@G$_{0}$W$_{0}$@PBE optical spectra.  
The dielectric function comprises a real part, Re($\varepsilon$), describing polarization and dielectric screening, and an imaginary part, Im($\varepsilon$), associated with optical absorption. The calculated dielectric spectra of Ba$_{3}$MA$_{3}$ (A = Cl, Br) and Ba$_{3}$MI$_{3}$ are shown in Figs.~S6 and \ref{fig:2}(e--h), respectively. The electronic dielectric constant ($\varepsilon_{\infty}$) increases from Cl- to Br- to I-based systems, indicating stronger screening and reduced carrier recombination in iodides. Correspondingly, the absorption onset and first peak ($E_o$) redshift with decreasing quasiparticle band gaps (Table~\ref{tab:1}). The resulting optical band gaps (0.97--1.81~eV) lie within the optimal range for efficient solar energy conversion.

In addition to optical properties, key excitonic parameters—including exciton binding energy ($E_{B}$), temperature ($T_{exc}$), radius ($r_{exc}$), and lifetime ($\tau_{exc}$)—are evaluated, as excitons formed upon photoexcitation strongly affect charge separation and recombination. A lower $E_{B}$ promotes efficient carrier dissociation and improved device performance. From first-principles calculations, $E_{B}$ is obtained as the difference between the direct quasiparticle band gap (G$_{0}$W$_{0}$@PBE+SOC) and the optical band gap (BSE@G$_{0}$W$_{0}$@PBE+SOC) \citep{chapter2-38,chapter5-18}. Table~\ref{tab:2} summarizes the calculated $E_{B}$ values, which decrease from Cl- to Br- to I-based compounds and lie in the range 0.254--0.352~eV. \textcolor{black}{These values are significantly larger than those of conventional 3D halide perovskites \citep{chapter5-14}, indicating pronounced excitonic effects that are beneficial for light-emitting devices, lasers, and certain photodetectors; however, their advantage in photovoltaics depends on efficient exciton dissociation.}

To further analyze excitonic behavior, the Wannier--Mott model \citep{chapter5-16} is employed. The effective dielectric constant ($\varepsilon_{\mathrm{eff}}$) lies between the electronic ($\varepsilon_{\infty}$) and static ($\varepsilon_{s}=\varepsilon_{\infty}+\varepsilon_{i}$) values, where the ionic contribution $\varepsilon_{i}$ is obtained from DFPT calculations (see Sec. XII of the SM \cite{supp}). For all compounds, $E_{B}\gg\hbar\omega_{LO}$, with $\omega_{LO}$ the longitudinal optical phonon frequency (Tables~\ref{tab:2} and \ref{tab:3}), indicating dominant electronic screening and $\varepsilon_{\mathrm{eff}}\approx\varepsilon_{\infty}$ \citep{chapter1-65,chapter1-66}. Using $E_{B}$, $\varepsilon_{\infty}$, and $\mu^{*}$, additional excitonic parameters, including $T_{exc}$, $r_{exc}$, and the zero-separation wave-function probability $|\phi_{n}(0)|^{2}$, are determined (Table~\ref{tab:2}). The calculated exciton radii (0.90--1.95~nm) exceed the lattice constants ($\sim$ 6--7~\AA), indicating intermediate-radius excitons extending over multiple unit cells and lying between Frenkel and Wannier--Mott limits \citep{AP-13,AP-14}.  

Moreover, the inverse of $|\phi_{n}(0)|^{2}$ provides a qualitative estimate of the exciton lifetime ($\tau_{exc}$), reflecting the recombination timescale (see Sec. XIII of the SM \cite{supp}). Notably, I-based compounds exhibit longer $\tau_{exc}$ than their Br- and Cl-based counterparts, implying reduced recombination rates and enhanced quantum yield and conversion efficiency.

In a recent study, Filip \textit{et al.} \citep{chapter3-39} introduced phonon screening into the exciton binding energy $E_{B}$ by considering four key parameters: the reduced effective mass ($\mu^{*}$), the electronic ($\varepsilon_{\infty}$) and static ($\varepsilon_{s}$) dielectric constants, and the longitudinal optical phonon frequency ($\omega_{LO}$). To obtain a representative $\omega_{LO}$ from multiple phonon branches, we adopted the method of Hellwarth and Biaggio \citep{chapter2-22}. Assuming isotropic and parabolic band dispersion, the phonon-screening correction to the exciton binding energy is given by

\begin{equation}
\Delta E_{B}^{ph}=-2\omega_{LO}\left(1-\frac{\varepsilon_{\infty}}{\varepsilon_{s}}\right)\frac{\sqrt{1+\omega_{LO}/E_{B}}+3}{\left(1+\sqrt{1+\omega_{LO}/E_{B}}\right)^{3}}.
\label{eq:1}
\end{equation}

\begin{table}[t]
\caption{\label{tab:2}Calculated excitonic parameters ($E_{B}$, $T_{exc}$, $r_{exc}$ and $|\phi_{n}(0)|^{2}$), phonon screening corrections
($\Delta E_{B}^{ph}$), and modified exciton binding energy ($E_{B}^{\prime}$)
of Ba$_{3}$MA$_{3}$ (M = P, As, Sb, Bi; A = Cl, Br, I).}

\centering{}%
\begin{tabular}{cccccccccc}
\hline 
\multirow{2}{*}{Systems} & \multirow{2}{*}{} & \multirow{2}{*}{A} & \multirow{2}{*}{} & $E_{B}$ & $T_{exc}$ & $r_{exc}$ & $|\phi_{n}(0)|^{2}$ & $\Delta E_{B}^{ph}$ & $E_{B}^{\prime}$\tabularnewline
 &  &  &  & (eV) & (K) & (nm) & (10$^{27}$m$^{-3}$) & (meV) & (eV)\tabularnewline
\hline 
\multirow{3}{*}{Ba$_{3}$PA$_{3}$} &  & Cl &  & 0.352 & 4081 & 0.97 & 0.35 & -10.18 & 0.342\tabularnewline
 &  & Br &  & 0.328 & 3803 & 1.04 & 0.28 & -8.57 & 0.319\tabularnewline
 &  & I &  & 0.272 & 3154 & 1.25 & 0.16 & -7.92 & 0.264\tabularnewline
 &  &  &  &  &  &  &  &  & \tabularnewline
\multirow{3}{*}{Ba$_{3}$AsA$_{3}$} &  & Cl &  & 0.341 & 3954 & 0.90 & 0.44 & -8.02 & 0.333\tabularnewline
 &  & Br &  & 0.320 & 3710 & 1.03 & 0.29 & -6.87 & 0.313\tabularnewline
 &  & I &  & 0.256 & 2968 & 1.38 & 0.12 & -6.06 & 0.250\tabularnewline
 &  &  &  &  &  &  &  &  & \tabularnewline
\multirow{3}{*}{Ba$_{3}$SbA$_{3}$} &  & Cl &  & 0.343 & 3977 & 1.03 & 0.29 & -6.25 & 0.337\tabularnewline
 &  & Br &  & 0.318 & 3687 & 1.09 & 0.25 & -6.09 & 0.312\tabularnewline
 &  & I &  & 0.271 & 3142 & 1.30 & 0.15 & -6.72 & 0.264\tabularnewline
 &  &  &  &  &  &  &  &  & \tabularnewline
\multirow{3}{*}{Ba$_{3}$BiA$_{3}$} &  & Cl &  & 0.307 & 3559 & 1.51 & 0.09 & -5.76 & 0.301\tabularnewline
 &  & Br &  & 0.282 & 3270 & 1.65 & 0.07 & -5.84 & 0.276\tabularnewline
 &  & I &  & 0.254 & 2945 & 1.95 & 0.04 & -6.83 & 0.247\tabularnewline
\hline 
\end{tabular}
\end{table}

Using Eq.~(\ref{eq:1}), we find that phonon screening reduces the exciton binding energy by only 1.82–2.91\%, yielding modified values ($E_{B}^{\prime}$) between 0.247 and 0.342~eV (see Table~\ref{tab:2}). This modest reduction indicates that ionic (phonon) contributions are negligible and that dielectric screening is dominated by the electronic component.

To assess intrinsic carrier mobility limits, we calculated the mobility using the simulated band structures \citep{chapter2-20,chapter2-21}. In polar semiconductors, carrier scattering near room temperature is primarily governed by longitudinal optical (LO) phonons \citep{chapter5-16,chapter5-18}, making polaron formation—arising from carrier-phonon interactions—essential for realistic mobility estimates. This interaction is described by the Fröhlich model \citep{chapter2-51}, which introduces a dimensionless coupling constant $\alpha$ \citep{chapter2-51,Ref.37}:

\begin{equation}
\alpha=\frac{1}{4\pi\varepsilon_{0}}\frac{1}{2}\Big(\frac{1}{\varepsilon_{\infty}}-\frac{1}{\varepsilon_{s}}\Big)\frac{e^{2}}{\hbar\omega_{LO}}\Big(\frac{2m^{\ast}\omega_{LO}}{\hbar}\Big)^{1/2}.
\label{eq:2}
\end{equation}

Typically, $\alpha>10$ indicates strong coupling, whereas $\alpha\ll1$ corresponds to weak coupling \citep{chapter2-20,AP-15}. Table~\ref{tab:3} display the calculated values lying in the range, 2.16–4.01, which place these materials in the intermediate coupling regime. Polaron formation lowers the quasiparticle energies of both electrons and holes, with the polaron energy given by \citep{chapter5-16,Ref.36}:

\begin{equation}
E_{p}=(-\alpha-0.0123\alpha^{2})\hbar\omega_{LO}.
\label{eq:3}
\end{equation}

Table~\ref{tab:3} also shows a comparison of the resulting quasiparticle gaps with exciton binding energies, confirming that bound excitons are energetically more stable than charge-separated polaronic states in all compounds.

Following Feynman’s treatment of the Fröhlich Hamiltonian \citep{chapter2-23,chapter8-10}, the polaron effective mass is expressed as

\begin{equation}
m_{p}=m^{\ast}\Big(1+\frac{\alpha}{6}+\frac{\alpha^{2}}{40}+...\Big).
\label{eq:4}
\end{equation}

As summarized in Table~\ref{tab:3}, electron and hole effective masses increase by 48–107\% upon polaron formation, consistent with intermediate carrier–lattice coupling. The corresponding upper bound of the polaron mobility is estimated using the Hellwarth model \citep{chapter2-22,chapter8-9}:

\begin{equation}
\mu_{p}=\frac{3\sqrt{\pi}e}{2\pi c\omega_{LO}m^{*}\alpha}\frac{\sinh(\beta/2)}{\beta^{5/2}}\frac{w^{3}}{v^{3}}\frac{1}{K(a,b)},
\label{eq:5}
\end{equation}

where $\beta=hc\omega_{LO}/k_{B}T$, $w$ and $v$ are temperature-dependent variational parameters, and $K(a,b)$ is a function of $\beta$, $w$,
and $v$ defined in Sec. XV of the SM \cite{supp}. The highest calculated mobilities are found to be 25.71~cm$^{2}$V$^{-1}$s$^{-1}$ for electrons and 75.09~cm$^{2}$V$^{-1}$s$^{-1}$ for holes (see Table~\ref{tab:3}), indicating that despite carrier–phonon coupling, these antiperovskite derivatives retain mobilities suitable for efficient optoelectronic devices.
\begin{figure}[t]
\centering
\begin{centering}
\includegraphics[width=0.5\textwidth,height=1\textheight,keepaspectratio]{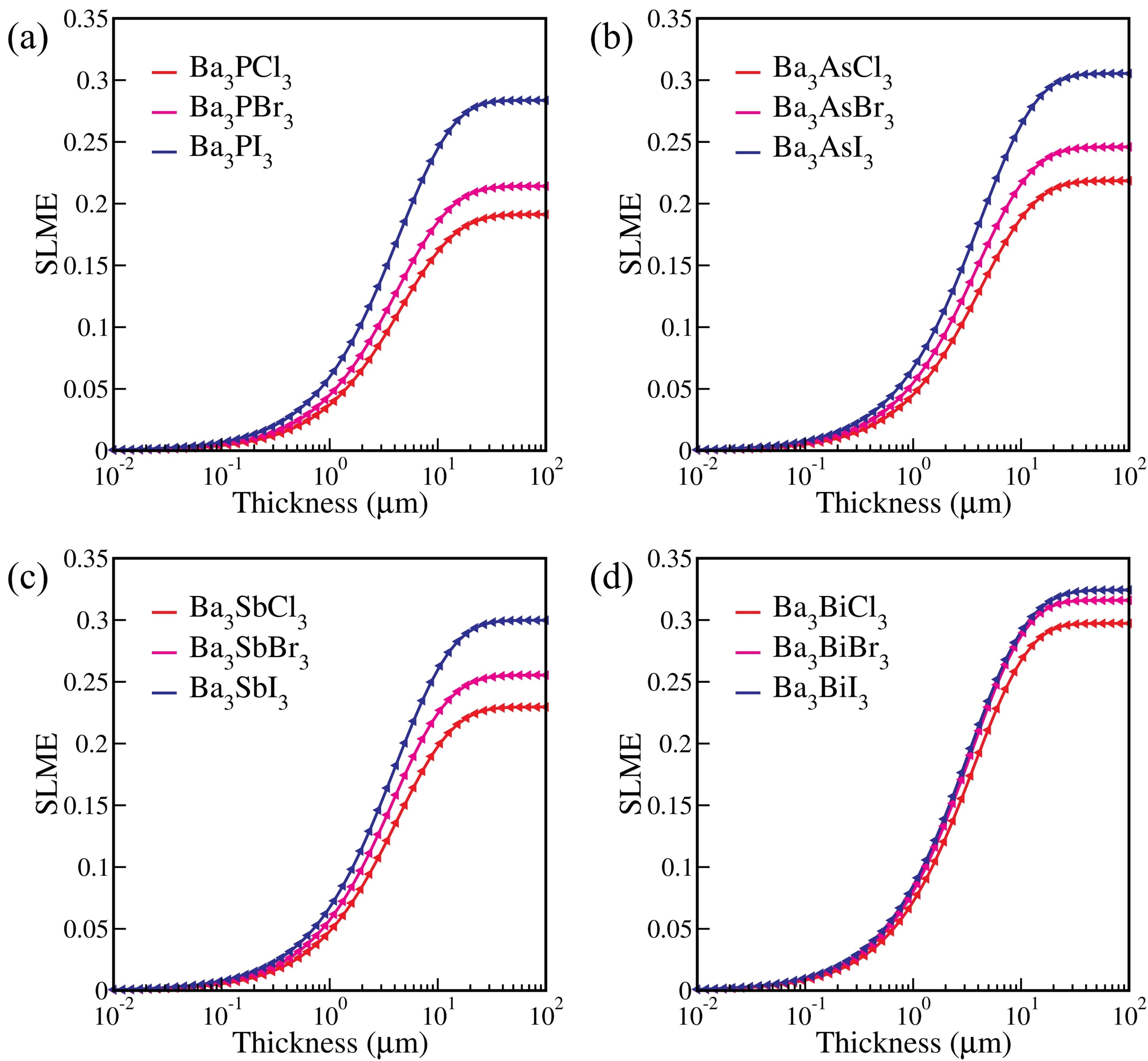}
\par\end{centering}
\caption{\label{fig:3}Spectroscopic limited maximum efficiency (SLME) of (a) Ba$_{3}$PA$_{3}$, (b) Ba$_{3}$AsA$_{3}$, (c) Ba$_{3}$SbA$_{3}$, and (d) Ba$_{3}$BiA$_{3}$ (A = Cl, Br, I), calculated using the BSE@G$_{0}$W$_{0}$@PBE+SOC method.}
\end{figure}

\begin{table*}[t]
\caption{\label{tab:3}Calculated polaron parameters and spectroscopic limited
maximum efficiency (SLME) of Ba$_{3}$MA$_{3}$ (M = P, As, Sb, Bi;
A = Cl, Br, I).}

\centering{}%
\begin{tabular}{cccccccccccccccccc}
\hline 
\multirow{2}{*}{Systems} & \multirow{2}{*}{} & \multirow{2}{*}{A} & \multirow{2}{*}{} & \multirow{2}{*}{$\omega_{LO}$ (THz)} & \multicolumn{2}{c}{$\alpha$} &  & \multicolumn{2}{c}{$E_{p}$ (meV)} &  & \multicolumn{2}{c}{$m_{p}/m^{*}$} &  & \multicolumn{2}{c}{$\mu_{p}$ (cm$^{2}$V$^{-1}$s$^{-1}$)} &  & \multirow{2}{*}{SLME (\%)}\tabularnewline
\cline{6-7} \cline{7-7} \cline{9-10} \cline{10-10} \cline{12-13} \cline{13-13} \cline{15-16} \cline{16-16} 
 &  &  &  &  & $e$ & $h$ &  & $e$ & $h$ & \multirow{1}{*}{} & $e$ & $h$ &  & $e$ & $h$ &  & \tabularnewline
\hline 
\multirow{3}{*}{Ba$_{3}$PA$_{3}$} &  & Cl &  & 3.87 & 3.54 & 2.60 &  & 59.20 & 43.00 &  & 1.90 & 1.60 &  & 9.58 & 28.62 &  & 19.10\tabularnewline
 &  & Br &  & 3.14 & 3.50 & 2.93 &  & 47.47 & 39.47 &  & 1.89 & 1.70 &  & 13.17 & 24.57 &  & 21.40\tabularnewline
 &  & I &  & 2.33 & 3.78 & 3.95 &  & 38.17 & 39.97 &  & 1.99 & 2.05 &  & 18.59 & 15.86 &  & 28.36\tabularnewline
 &  &  &  &  &  &  &  &  &  &  &  &  &  &  &  &  & \tabularnewline
\multirow{3}{*}{Ba$_{3}$AsA$_{3}$} &  & Cl &  & 3.14 & 3.80 & 2.87 &  & 51.72 & 38.64 &  & 1.99 & 1.69 &  & 9.07 & 24.44 &  & 21.84\tabularnewline
 &  & Br &  & 2.73 & 3.42 & 2.82 &  & 40.29 & 32.99 &  & 1.86 & 1.67 &  & 13.73 & 26.92 &  & 24.58\tabularnewline
 &  & I &  & 2.25 & 3.00 & 2.84 &  & 28.98 & 27.39 &  & 1.72 & 1.68 &  & 25.71 & 30.80 &  & 30.52\tabularnewline
 &  &  &  &  &  &  &  &  &  &  &  &  &  &  &  &  & \tabularnewline
\multirow{3}{*}{Ba$_{3}$SbA$_{3}$} &  & Cl &  & 2.51 & 3.99 & 2.88 &  & 43.51 & 31.00 &  & 2.06 & 1.69 &  & 10.65 & 33.26 &  & 22.94\tabularnewline
 &  & Br &  & 2.46 & 3.80 & 2.72 &  & 40.52 & 28.64 &  & 1.99 & 1.64 &  & 11.33 & 36.11 &  & 25.52\tabularnewline
 &  & I &  & 2.28 & 3.85 & 3.04 &  & 38.07 & 29.78 &  & 2.01 & 1.74 &  & 14.28 & 32.49 &  & 29.97\tabularnewline
 &  &  &  &  &  &  &  &  &  &  &  &  &  &  &  &  & \tabularnewline
\multirow{3}{*}{Ba$_{3}$BiA$_{3}$} &  & Cl &  & 2.44 & 3.16 & 2.21 &  & 33.17 & 22.94 &  & 1.78 & 1.49 &  & 20.52 & 68.06 &  & 29.73\tabularnewline 
 &  & Br &  & 2.36 & 3.13 & 2.16 &  & 31.77 & 21.67 &  & 1.77 & 1.48 &  & 21.00 & 73.10 &  & 31.60\tabularnewline 
 &  & I &  & 2.14 & 4.01 & 2.43 &  & 37.29 & 22.18 &  & 2.07 & 1.55 &  & 13.62 & 75.09 &  & 32.41\tabularnewline
\hline 
\end{tabular}
\end{table*}

To further evaluate optoelectronic performance, we calculated the spectroscopic limited maximum efficiency (SLME) \citep{chapter3-30}, which provides a more realistic assessment than the Shockley–Queisser (SQ) limit \citep{chapter3-31}. Before computing SLME, we examined whether optical transitions from the VBM to the CBM are allowed, as direct bandgap materials may still exhibit forbidden transitions due to inversion symmetry and identical band parities \citep{chapter5-16}. For this purpose, the transition dipole moment matrix element ($P$) was evaluated, where $P^{2}$ represents the transition probability. As shown in Figs.~\ref{fig:2} and S5, all compounds display allowed dipole transitions at the $\Gamma$ point.

Figure \ref{fig:3} shows the thickness-dependent SLME, computed using the BSE@G$_{0}$W$_{0}$@PBE+SOC approach. It increases with thickness before saturating. A clear trend is observed, with SLME rising from Cl- to Br- to I-based compounds, reflecting the more favorable band gaps of iodide systems. The maximum SLME values range from 19.10\% to 32.41\%. Notably, Ba$_{3}$AsI$_{3}$, Ba$_{3}$SbI$_{3}$, and Ba$_{3}$BiA$_{3}$ (A = Cl, Br, I) exhibit SLME values exceeding those of conventional lead-based perovskites, such as CsPbI$_{3}$ (20.67\% \citep{chapter5-17}) and MAPbI$_{3}$ (28.97\% \citep{chapter5-13}). \textcolor{black}{In addition, several other key device parameters have been computed within the SLME framework (see Sec. XVI of the SM \cite{supp})}. These results underscore the strong efficiency potential of antiperovskite derivatives for next-generation optoelectronic applications.

{\it Conclusion:}
In summary, we have presented a comprehensive first-principles investigation of the structural, excitonic, polaronic, and optoelectronic properties of cubic antiperovskite derivatives Ba$_3$MA$_3$ (M = P, As, Sb, Bi; A = Cl, Br, I). Our results establish these compounds as dynamically, thermodynamically, and mechanically stable direct-gap semiconductors with bandgaps spanning the optimal range for photovoltaic and optoelectronic applications. Explicit many-body perturbation theory calculations reveal moderate exciton binding energies and intermediate-radius excitons, indicating strong Coulomb interactions while still enabling efficient charge separation. Carrier-phonon coupling is found to be in the intermediate regime, leading to polaron formation with mobilities that remain favorable for device operation.
Importantly, spectroscopic limited maximum efficiency calculations predict power conversion efficiencies of up to 32\%, exceeding those of several state-of-the-art lead-based perovskites. These findings highlight the crucial role of excitonic and polaronic effects in governing charge transport and optical response in antiperovskite derivatives. Overall, this work identifies Ba-based antiperovskite derivatives as a robust, lead-free materials platform and provides key physical insights to guide the design of next-generation high-efficiency optoelectronic devices.

{\it Acknowledgments:}
S.A would like to acknowledge Indian Institute of Technology Bombay, India for financial support. P.J would like to acknowledge the Council of Scientific and
Industrial Research (CSIR), Government of India {[}Grant No. 3WS(007)/2023-24/EMR-II/ASPIRE{]}
for financial support. The authors acknowledge the High Performance Computing Cluster (HPCC)
\textquoteleft Magus\textquoteright{} at Shiv Nadar Institution of
Eminence for providing computational resources that have contributed
to the research results reported within this paper.

%{\it DATA AVAILABILITY:}
%The data that support the findings of this article are not publicly
%available. The data are available from the authors upon reasonable
%request.

% \bibliographystyle{apsrev4-2}

\bibliography{refs}

\end{document}